\documentclass{aa}  

\usepackage{graphicx}
\usepackage{txfonts}
\usepackage{appendix}
\usepackage{mathrsfs}
\usepackage{isotope}
\usepackage{color}

\begin{document} 
\newcommand{\mic} {$\mu$m}
\newcommand{\rw}{RW Aur A\,}
\newcommand{\HI}{\ion{H}{i}}
\newcommand{\brg}{Br$\gamma$}
\newcommand{\NaI}{\ion{Na}{i}}
\newcommand{\macc}{$\dot{M}_{acc}$}
\newcommand{\lacc}{L$_{acc}$}
\newcommand{\lbol}{L$_{bol}$}
\newcommand{\kms}{km\,s$^{-1}$}
\newcommand{\um}{$\mu$m}
\newcommand{\lam}{$\lambda$}
\newcommand{\msyr}{M$_{\odot}$\,yr$^{-1}$}
\newcommand{\Av}{A$_V$}
\newcommand{\msun}{M$_{\odot}$}
\newcommand{\Lsun}{L$_{\odot}$}
\newcommand{\Lstar}{L$_{\star}$}
\newcommand{\Tstar}{T$_{\star}$} 
\newcommand{\Nco}{$N_{CO}$} 
\newcommand {\Tbb} {$T_{\rm BB}$}
\newcommand{\up}{$\upsilon$}
\newcommand{\egs}{erg\,s$^{-1}$\,cm$^{-2}$\,\AA$^{-1}$}
\newcommand{\simless}{\mathbin{\lower 3 pt\hbox {$\rlap{\raise 5pt\hbox{$\char'074$}}\mathchar"7218$}}} 
\newcommand{\simgreat}{\mathbin{\lower 3pt\hbox {$\rlap{\raise 5pt\hbox{$\char'076$}}\mathchar"7218$}}} 

   \title{Exploring the dimming event of RW Aur A through multi-epoch VLT/X-Shooter spectroscopy
    \thanks{Based on observations collected at the European Southern Observatory Paranal, Chile (ESO programme 294.C-5047 and 098.C-0922(A)).}}
   \subtitle{}

   \author{M. Koutoulaki
          \inst{1,2}
          \and
          S. Facchini\inst{3}
          \and
          C. F. Manara\inst{3}
          \and
          A. Natta\inst{1,4}
          \and
          R. Garcia Lopez\inst{1}
          \and
          R. Fedriani\inst{1,2}
          \and
          A. Caratti o Garatti\inst{1}
          \and
          D. Coffey\inst{2,1}
          \and
          T. P. Ray\inst{1}
          }

  \institute{Dublin Institute for Advanced Studies, 31 Fitzwilliam Place, Dublin 2, Ireland\\
              \email{mariakout@cp.dias.ie}\\
         \and
             School of Physics, University College Dublin, Belfield, Dublin 4, Ireland\\
                  \and
                     European Southern Observatory, Karl-Schwarzschild-Strasse 2, Garching bei München, 85748 Germany\\
                  \and
                     INAF/Osservatorio Astrofisico di Arcetri, Largo E. Fermi 5, 50125 Firenze, Italy\\
                    }

   \date{}

 
  \abstract
   {\rw is a classical T Tauri star that has suddenly undergone three major dimming events since 2010. The reason for these dimming events is still not clear.}
   {We aim to understand the dimming properties, examine accretion variability, and derive the physical properties of the inner disc traced by the CO ro-vibrational emission at near-infrared wavelengths (2.3 \um). }
   {We compared two epochs of X-Shooter observations, during and after the dimming. We modelled the rarely detected CO bandhead emission in both epochs to examine whether the inner disc properties had changed. The spectral energy distribution was used to derive the extinction properties of the dimmed spectrum and compare the infrared excess between the two epochs. Lines tracing accretion were used to derive the mass accretion rate in both states.}
   {The CO originates from a region with physical properties of T=3000\,K, N$_{\rm CO}$=1$\times$10$^{21}$\,cm$^{-2}$ and $\rm v_{k}\sin{i}$=113 km/s. The extinction properties of the dimming layer were derived with the effective optical depth ranging from $\tau_{eff}$ $\sim$2.5-1.5 from the UV to the near-IR. The inferred mass accretion rate $\mathrm{\dot{M}_{acc}}$ is $\sim 1.5\times 10^{-8}$ \msun/yr and $\sim 2\times 10^{-8}$ \msun/yr after and during the dimming respectively. By fitting the spectral energy distribution, additional emission is observed in the infrared during the dimming event from dust grains with temperatures of 500-700\,K.}
   { The physical conditions traced by the CO are similar for both epochs, indicating that the inner gaseous disc properties do not change during the dimming events. The extinction curve is flatter than that of the interstellar medium, and large grains of a few hundred microns are thus required. When we correct for the observed extinction, the mass accretion rate is constant in the two epochs, suggesting that the accretion is stable and therefore does not cause the dimming. The additional hot emission in the near-IR is located at about 0.5 au from the star and is not consistent with an occulting body located in the outer regions of the disc. The dimming events could be due to a dust-laden wind, a severe puffing up of the inner rim, or a perturbation caused by the recent star-disc encounter.}

   \keywords{protoplanetary disks --
             stars: formation --
             stars: individual: RW Aurigae --
             stars: variables: T Tauri, Herbig Ae/Be --
             ISM: dust, extinction --
             accretion, accretion disks
               }

   \maketitle
%

\section{Introduction}

The binary RW Aur system comprises two Classical T Tauri stars (CTTSs), \rw and RW Aur B (see Table \ref{tab:param} for properties of the primary star \rw). The system is located at a distance of 152 pc \citep{GAIA2016,GAIA2018}. The primary component has a spectral type between K1-K4 \citep{Petrov2001} and $\mathrm{V_{mag}\sim 10.5}$ \citep{White2001}. It has a dusty disc radius of $\leq 57$ au, making it one of the few T Tauri stars with such a small disc \citep{Cabrit2006,Rodriguez2018}. It is also one of the very few T Tauri stars that have a jet \citep{Dougados2000,Lopez2003,Beck2008}. This system has captured attention mainly through its dimming events, which suddenly started in 2010 \citep{Rodriguez2013,Petrov2015}. During the first dimming event, the primary star dropped by $\sim$1.5-2 mag in the V band, and the occultation lasted almost 180 days \citep{Rodriguez2013}. The second dimming event was even more significant, as the dimming was more than two magnitudes \citep{Petrov2015,Bozhinova2016}. 

\begin{table}[ht]
\caption{Stellar properties of \rw }             
\label{tab:param}      
\centering                          
\begin{tabular}{l l c}        
\hline\hline                 
Parameters & Value & Reference\\    
\hline  
   RA (2000)& 05 07 49.538 &\\
   DEC(2000) & +30 24 05.07 &\\
   spectral type & K1$\pm2$  &1\\
   K (mag) & 7.06$\pm$0.17  &1\\
   mass & 1.3-1.4 $M_{\odot}$  &2, 3, 6\\      
   d & 152 pc &4\\  
   log$L_{*}$ &$0.23\pm0.14\, L_{\odot}$  &1\\
  $R_{*}$ &$1.1\, R_{\odot}$ & 5\\
  T&  5080\,K  &1 \\
   
\hline                                   
\end{tabular} 
\tablefoot{(1)\, \citet{White2001}, (2) \,Ghez et al. 1997b, (3)\, Woitas et al. 2001, (4)\, \citet{GAIA2018},  (5)\, \citet{Ingleby2013}, (5)\, \citet{Rodriguez2018}}
\end{table}

The dimming events of RW Aur are similar (1-3 mag in depth) to those of UX Ori stars \citep{Grinin1991}, which are typically associated with earlier type objects. A few other T Tauri stars are known to undergo similar dimming events, for instance, AA Tau \citep{Bouvier2013} and V409 Tau \citep{Rodriguez2015}. The origin of these dimming events is still debated, ranging from inner disc warps to puffed-up inner rims. RW Aur is a unique case that may shed light on the origin of this phenomenon, because of the large amount of data that is available before and during the dimming events, with both photometric and spectroscopic monitoring. In particular, for \rw the occulting material has been associated with the inner regions of the disc due to additional emission (excess) at near-IR (NIR) wavelengths and more particularly in L and M bands \citep{Shennavrin2015,Bozhinova2016}. A large ($\sim$600 au) tidal arm in \isotope[12]{CO} was detected by \citet{Cabrit2006}. This arm was reproduced by \citet{Dai2015} through modelling the dynamical encounter of the two stars. Comparing X-ray observations in the bright and dim states, \citet{Schneider2015} found an increase in the gas column density during the second dimming event. Additional X-ray observations by \citet{gunther2018} also show an increased gas column density as well as an iron enhancement in the stellar corona. They studied the extinction from optical \citep{Antipin2015} to NIR \citep{Schneider2015}, and found that it is nearly grey, suggesting grain growth in the inner disc. Other authors \citep[see e.g.][]{Petrov2015} suggested that the extinction could be due to an outburst or a wind. \citet{Facchini2016}, on the other hand, argued that these dimming events are due to a perturbation of a misaligned or warped inner disc.

Interestingly, \rw is one of the few T Tauri stars known to date to show the CO bandhead in emission \citep{Carr1989,Beck2008,Eisner2014}. These transitions between the CO vibrational states trace the innermost regions of the disc, and changes in the line properties are expected if the inner gaseous disc varies significantly across the dimming events. In this paper, we present spectroscopic observations of \rw obtained with the X-SHooter on the Very Large Telescope (VLT/X-Shooter), ranging from UV to NIR wavelengths, with a particular focus on the properties of the CO bandhead emission. We also aim to probe whether the dimming events have been caused by a significant variability in the mass accretion rate. Finally, we infer the properties of the dust within the occulting material from the extinction curve across the dimming events. The observations and the data reduction are discussed in Section 2, and the results and discussions in Sections 3, 4, and 5, respectively. In Section 6 we summarise the results and draw our conclusions.

\section{Observations and data reduction}
\rw was observed using the VLT/X-Shooter instrument \citep{Vernet2011} on 2015 March 19 (DDT Pr.Id. 294.C-5047, PI Facchini) and on 2016 September 30 (Pr.Id. 098.C-0922(A), PI Facchini). The wavelength coverage is from 0.3 to 2.4 \mic. 
For both observing runs, a set of large (1.6 $\times$11$\arcsec$-1.5 $\times$11$\arcsec$-1.2 $\times$11$\arcsec$) and narrow (0.5 $\times$11$\arcsec$-0.4 $\times$11$\arcsec$-0.4 $\times$11$\arcsec$) slits were used in the UVB, VIS, and NIR arms, respectively. The seeing at the time of the observations was 1.00 for both epochs. The position angle in the 2016 run was 178$\degr$, and in the 2015 run it was -156$\degr$ (east of north).  While the data taken with the large slits have lower spectral resolution, they do not have any slit losses, thus they were flux calibrated and were used to calibrate the narrow slit data. In all the observations, care has been taken in order to avoid any contribution from RW Aur B (separation of $\sim$1.5$\arcsec$). The airmass for the observations was 1.7 and 2 for the 2016 and 2015 observations respectively. The data were reduced using the X-Shooter pipeline version 2.8.4. The spectra were flux calibrated using standard photometric stars observed during the night. They were corrected for telluric absorption using the telluric standards Hip045742 and Hip014898 for the 2016 and 2015 observations, respectively. The resolution of the observations in the NIR was computed to be R=7000 ($\Delta \rm v$=43\,km s$^{-1}$) from absorption telluric features in the spectrum. 

\begin{figure*}[ht]
\centering
\includegraphics[scale=0.4]{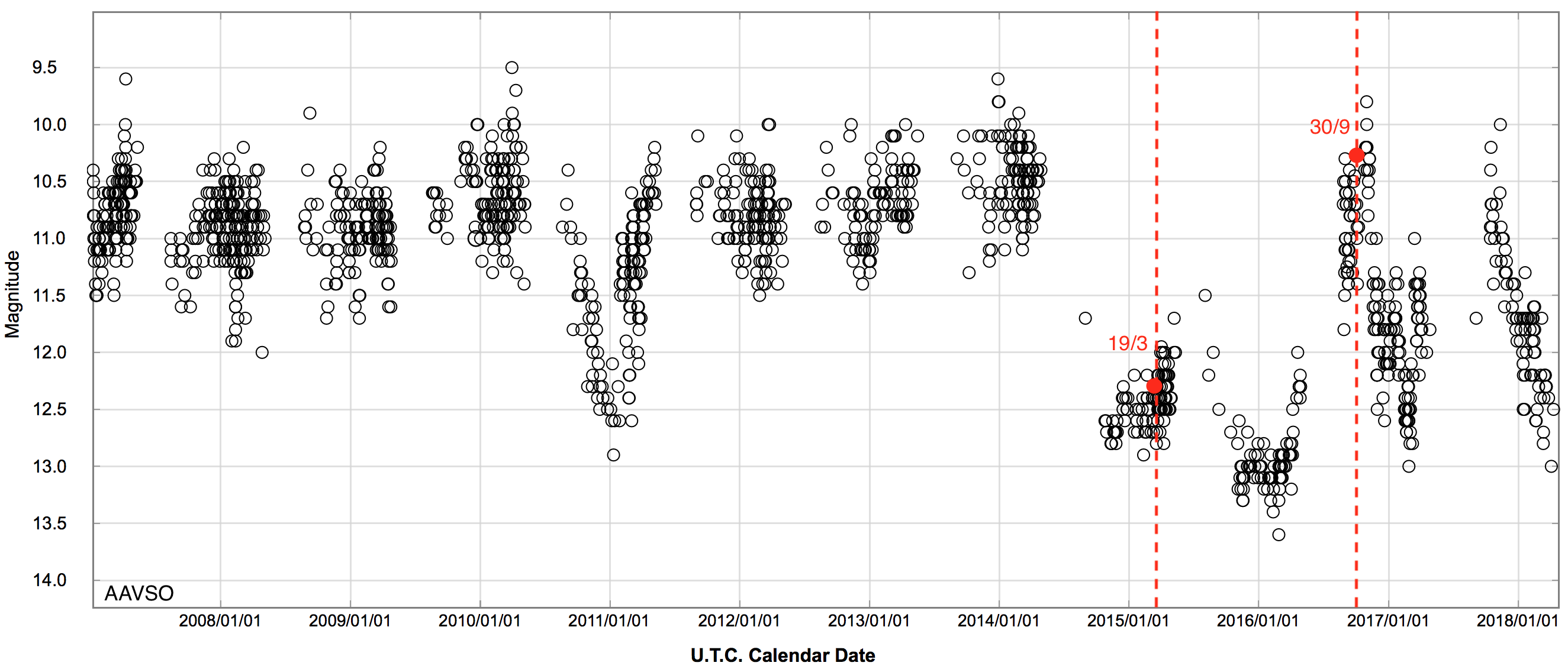}
\caption{Light curve of the RW Aur system in the visual. The plot is taken from The American Association of Variable Star Observers (AAVSO). The red dots indicate the time of our observations. The magnitudes for the two epochs were computed by producing synthetic photometry from the spectra.}
\label{fig:light_curve}
\end{figure*}
\section{Results}
\subsection{Spectra}
\label{spectra}

\begin{figure*}[ht]
\centering
\includegraphics[scale=0.8]{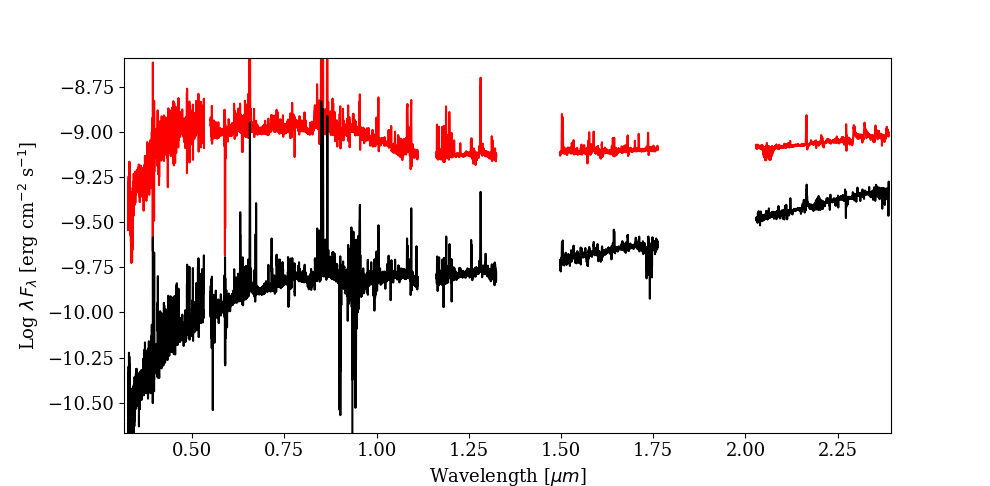}
\caption{Whole X-Shooter spectrum for the bright (red) and the dim (black) state in the wavelength range of 0.3-2.4 \mic. The y-axis is in logarithmic scale. A zoom-in of the spectra is shown in the appendix.}
\label{fig:spectra}
\end{figure*}

Figure \ref{fig:light_curve} shows the light curve of RW Aur during 2007--2018. On 2015 March 19, when the first spectrum was obtained, the star was in a deep minimum, while in 2016 September 30, it was back to its typical brightness. In the following, we refer to the first as the dim state and to the second as the bright state. The two X-Shooter spectra are shown in Figure \ref{fig:spectra}. The dim spectrum is weaker than the bright spectrum by a factor of 11.8 in the B band and up to 2.1 in the K band.

 Both spectra are rich in strong emission lines, making \rw an  unusual CTTS \citep[e.g.][]{Joy1945}. These lines are quite complex and receive contributions from multiple components.
 
 In both states, there is excess emission at short wavelengths, but there is no evidence of the Balmer jump (0.364 \mic) that is often seen in many other CTTS \citep[e.g][]{Alcala2014,Alcala2017}. A zoom-in in the spectra is shown in the appendix.


\subsection{CO overtone emission}
\label{sec:overtone}

\begin{figure*}[ht]
\centering\includegraphics[scale=0.7]{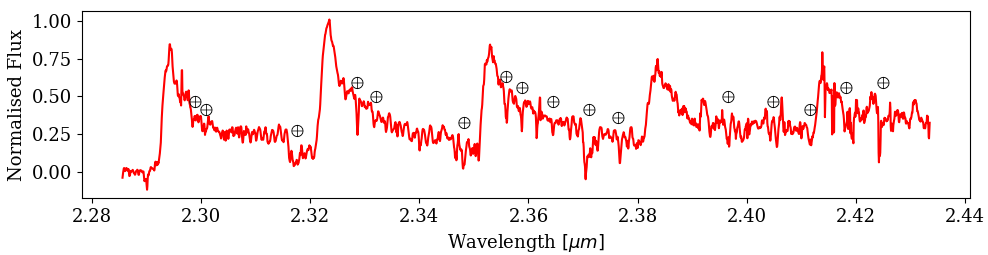}
\caption{CO continuum-subtracted spectrum of \rw for the bright state. The spectrum has been normalised to the second bandhead. The Earth symbol shows regions with strong telluric features.}
\label{fig:co_spec}
\end{figure*}

\rw is one of the very few CTTS that exhibit NIR CO overtone emission in the 2.3 \mic\ region. This emission is detected in both epochs, and the bright state is shown in Fig. \ref{fig:co_spec}. 
Both spectra are continuum-subtracted and normalised to the peak intensity of the second bandhead. 

In the bright state, we detect five ro-vibrational transitions ($\upsilon\mathrm{=2-0, 3-1, 4-2, 5-3,\, and\,6-4}$). The overall profile of the first four bandheads is clearly visible, with a sharp peak and  a small but significant "left shoulder", probably due to Keplerian broadening \citep[e.g.][]{Chandler1995}. The second bandhead peak ($\upsilon$ = 3-1) is slightly brighter than the first ($\upsilon$ = 2-0). At the spectral resolution of the observations,  we spectrally resolve several individual \textit{J} components of the rotational ladder. This is more obvious in the first two bandheads, as is clearly visible in Fig. \ref{fig:co_spec}. 

The CO spectrum of the dim state along with the model, which will be explained later on, is shown in Fig. \ref{fig:dim_model}. The data are convolved to a lower resolution (R$\sim$ 3500). 
As before, the data are normalised to the peak of the second bandhead and are shown in black. The dim spectrum is more noisy, and the correction for the telluric absorption much more uncertain. However, we clearly detect the first two bandheads and have a convincing signature of emission in the third, even if its profile is poorly retrieved.


\subsubsection{Modelling the CO emission}
\label{sec:model}
The CO spectrum provides very interesting information on the physical properties of the emitting region. We computed synthetic spectra using the  method outlined by \citet{Kraus2000}, which is briefly discussed below. \\
The energy of a diatomic molecule of a rotational level $J$ and vibrational level $\mathrm{\upsilon}$ is given by \citet{Dunham1932}. 
We assume that the levels are  populated in  local thermodynamic equilibrium (LTE).


The CO emitting region is described by  a single ring of width $\Delta r$, temperature ($\textit{T}$), and column density ($N_{\rm CO}$),  in  Keplerian rotation  around the central star. 
The $\upsilon$ band consists of 100 rotational transitions, and each \textit{J} component has a Gaussian profile with  width $\Delta \rm v$.\,
We created a spectrum over the range covered by the observed bandheads, with very high spectral resolution, so that each individual \textit{J} transition profile was well sampled, and computed the total optical depth at each wavelength. The intensity is then given as $I(\upsilon, \textit{J})=B_{\nu}(T)(1-e^{-\tau_{\upsilon,\textit{J}}})$. 
The resulting spectrum was then convolved over the Keplerian velocity pattern and reduced to the observed spectral resolution.\\

The models have four free parameters: temperature (\textit{T}), column density of the CO ($N_{\rm CO}$), intrinsic line width ($\Delta\nu$), and projected Keplerian velocity ($\rm v_{k} \sin{i}$), where $i$ is the inclination with respect to the line of sight ($\textit{i}=0$ for face-on ring).

\subsubsection{Physical conditions of the CO emitting region in the bright state}
\label{Results:CO conditions}
We have computed a large  grid of models by varying the free parameters over a broad range of values to compare the results to the CO bright state spectrum.
The projected Keplerian velocity is quite well constrained even though our spectra do not have very high spectral resolution.  As mentioned in Section \ref{sec:overtone}, a number of  components in the tail of the bands are spectrally resolved.  The rotational ladder, along with the blue shoulder of the bandhead that is reproduced by the model of Keplerian rotation, constrains the range of possible $\rm v_{k} \sin{i}$. At our spectral resolution, we find that what appears in the spectrum as a single $J$ component is, in fact, the overlap of the red-shifted and blue-shifted components of two adjacent $J$ transitions. Therefore, the peak position of the lines observed at the tail of the bandhead  does not coincide with the wavelength of any individual $J$ component, but  depends on $\rm v_{k} \sin{i}$ (see Fig.\ref{fig:jcomp}). 
The best representative value of $v_{\rm k} \sin(i)$, reproducing the wavelength of the best measured individual features in the first and second bandhead tails, is   113\,\kms. 
We can derive the value of $v_{\rm k}$ if the inclination is know.
From the literature, there are two values for the inclination of the disc. From ALMA studies, it has been confirmed that the inclination of the outer disc is around $50^{\circ}$ \citep{Rodriguez2018}. From KECK interferometric observations of a sample of young stellar objects (YSOs) by \citet{Eisner2014}, they estimated an inclination of the inner gaseous disc of \rw of $75^{\circ}$. 
 Table \ref{tab:best_fit} gives the values of $v_{\rm k}$ for the two inclinations. 

Temperature and column density determine the shape of the bandheads and the relative intensity of the different bands.
We find that we need a relatively warm gas of temperature $T\sim$3000\,K in order to reproduce the observed profiles. The gas is quite dense, with a CO column density $N$$\mathrm{_{CO}\approx 1\times10^{21}\,cm^{-2}}$. While the emission of the first three bandheads is mainly optically thick, the higher $\upsilon$ transitions, namely the fourth and fifth, still remain optically thin. Thus, an increase of the optical depth is associated with a slight change in the \textit{J} components of the tail of the first three bandheads. In addition, a significant increase in the intensity of the forth and the fifth bandhead would be expected.  We give the range of our estimates for $T$ and $N_{\rm CO}$ in  Table \ref{tab:best_fit}.  

The intensity of the bandheads, once $T$ and $N_{\mathrm{CO}}$ are known, depends only on the area of the emitting region. In the bright state, we find a value of about 2.5--3.1$\times$10$^{23}$\,cm$^{2}$, where the uncertainty depends mostly on the uncertainty on $T$. 


\subsubsection{CO spectrum in the dim state}
As the dim spectrum is much noisier, we used the fit of the bright state to compare it with the spectrum. In Fig. \ref{fig:dim_model} the dim state spectrum is shown along with the best representative model of the bright state. The model reproduces the observed spectrum well within the errors. The physical conditions of the CO emitting region therefore do not change significantly between the bright and dim states.


\begin{table*}[ht]
\caption{Inner disc properties}             
\label{tab:best_fit}      
\centering                          
\begin{tabular}{c c c c c c c }        
\hline\hline                 
Temperature &CO column density&$\Delta\rm v$&$\rm v_{k}\sin{i}$&\multicolumn{2}{c}{$\rm v_{k}$} \\ 
K& $\mathrm{cm^{-2}}$&\kms&\kms&\kms\\
&&&&\textit{i}=50$^{\circ}$&\textit{i}=70$^{\circ}$\\
\hline                        
    $3000^{+200}_{-300}$&$(1^{+1}_{-0.3})\times 10^{21}$&20&113&156&120\\      
   \hline                                   
\end{tabular} 
\end{table*}
\begin{figure*}[h!]
\centering
\includegraphics[scale=0.8]{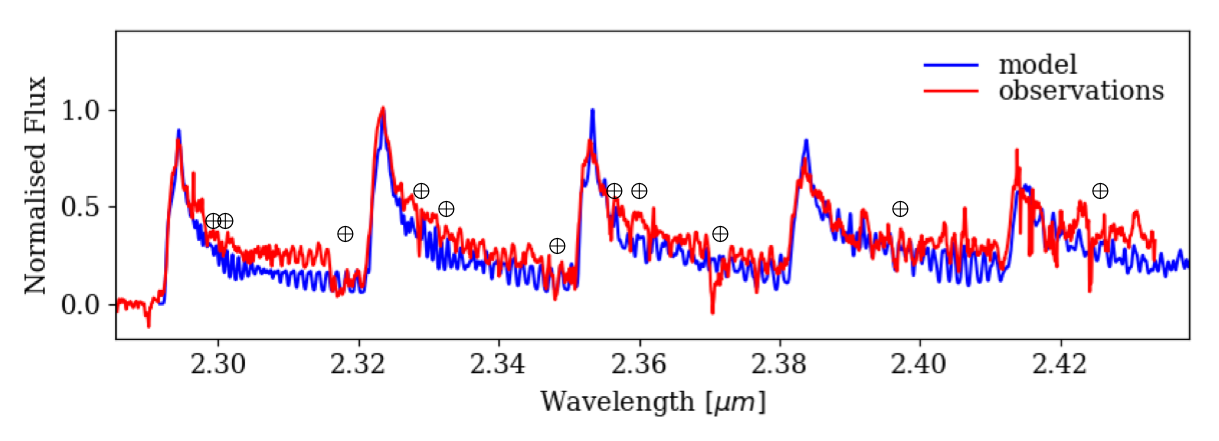}
\caption{Representative model and data of the bright state of the spectrum. Blue corresponds to the model and red to the observations.  The intensity is normalised to the second bandhead and the wavelength is in microns. Five bandheads are shown ($\mathrm{\upsilon=2-0, 3-1, 4-2, 5-3,\,and\,6-4}$). The parameters of the model are listed in Table \ref{tab:best_fit}. This particular case is for T=3000\,K, N$_{\mathrm{CO}}$=$8\times10^{20}\,$cm$^{-2}$, and $\rm v_{k}\sin{i}=$113\,\kms. }
\label{fig:best_fit_bright_bheads}
\end{figure*}
\begin{figure}
\centering
\includegraphics[width=\columnwidth]{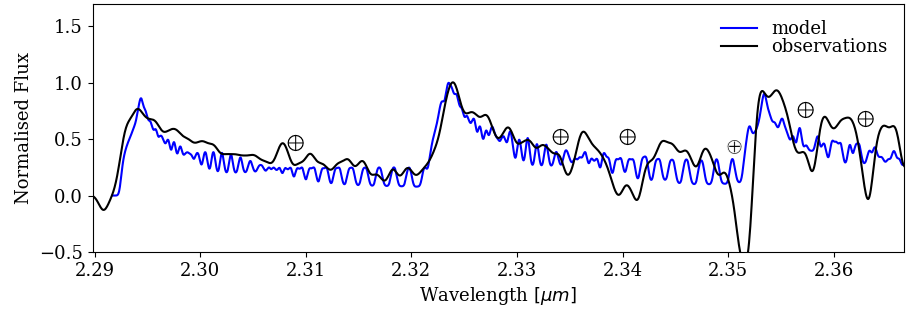}
\caption{First three bandheads of the dim state in black. In blue we plot the best-fit model of the bright state as in Fig. \ref{fig:best_fit_bright_bheads}.}
\label{fig:dim_model}
\end{figure}
\begin{figure}[ht!]
\centering
\includegraphics[width=\columnwidth]{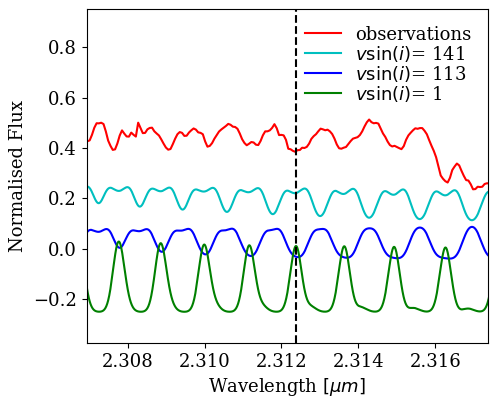}
\caption{\textit{J} components for the first bandhead for different $\rm v_{k}\sin{i}$. The spectra are shifted to facilitate comparison. The dashed vertical line corresponds to the intrinsic wavelength of one of the \textit{J} components.}
\label{fig:jcomp}
\end{figure}
\subsection{Extinction due to the dimming layer}
\label{extinction}
\begin{figure}[ht!]
\centering
\includegraphics[width=\columnwidth]{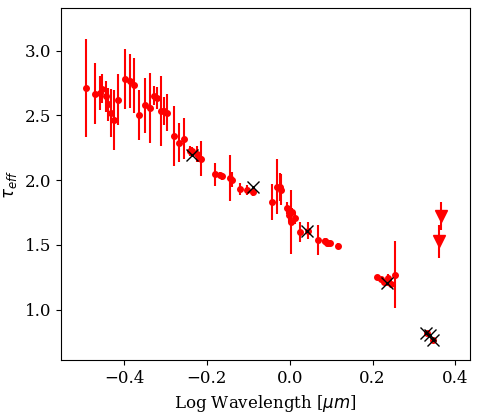}


\includegraphics[width=\columnwidth]{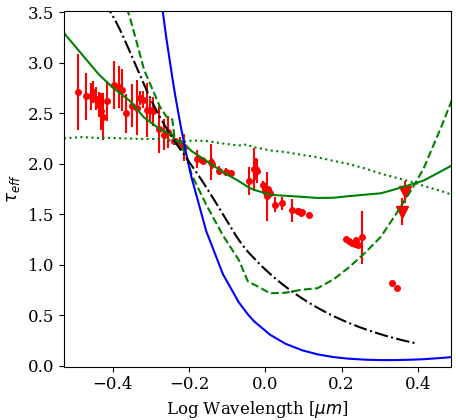}


\caption { Top panel: Effective optical depth of the dimming layer vs. wavelength computed from the ratio of the dim to the bright X-Shooter spectra.  The red circles  are the median of the  values over intervals of  5\,\AA, the error is the standard deviation within the interval. The triangles show the optical depth derived from the ratio of the CO bandheads, computed as the median of three points around the peak of  the first two bandheads. The black crosses correspond to continuum points in the spectrum. Bottom panel: The  red points (circles and triangles) are the same as in the top panel. The black dashed  line is the Cardelli extinction law for $R_{\rm V}=3.1$ \citep{Cardelli1989}. The green  curves plot $Q_{\rm abs}$ (dashed), $Q_{\rm scat}$ (dotted), and $Q_{\rm tot}=Q_{\rm abs}+0.12Q_{\rm scat}$ (solid) for a distribution of olivine  with minimum and maximum size $a_{\rm min}= 0.1$ \mic, $a_{\rm max}=150$ \mic, $q=3$ (see text). The blue curve plot $Q_{\rm tot}=Q_{\rm abs}+Q_{\rm scat}$ (solid)  for the same material, $a_{\rm min}=0.06$\mic, $a_{\rm max}=0.1$\mic, $q=3$. All curves are normalised to the observed optical depth at 0.58 \mic.}

\label{fig:ext}
\end{figure}

In this section we explore the possibility that the difference between the bright and dim state spectrum is due to a change in extinction, as previously suggested \citep[e.g][]{Petrov2015,Schneider2015,Rodriguez2016,Lamzin2017,Dodin2019}.
The magnitude and wavelength dependence of the  extinction between the two states traced by the X-Shooter spectra can be obtained by dividing the dim by the bright spectrum. The result is shown in Fig.~\ref{fig:ext} (top panel),  
where the  red points mark the median of the observed values over intervals of  5\,\AA, excluding the brightest emission lines, and the error is the standard deviation within the interval. The black crosses correspond to regions where very few emission lines are present.
The triangles show the values derived from the ratio of the two CO bandheads seen in the two states , assuming that the difference of the CO bands is due to extinction. They are higher than the value derived from the continuum adjacent to the bandheads, shown by one of the black crosses in Fig. \ref{fig:ext}. This discrepancy is easily explained if there is additional NIR  continuum emission when the star is dimmed, as suggested by \citet{Shennavrin2015}. In the following, we assume that  this is indeed the case and that the true values of the effective optical depth ($\tau_{eff}$) are better measured by the ratio of the CO bandheads.

The dependence of $\tau_{eff}$ on wavelength is well described by a power law $\tau_{eff}\propto \lambda^{-0.5} $ for $\lambda < 1.5 \mu m$. This is much flatter than the extinction curve of the interstellar medium (ISM), also shown in Fig.~\ref{fig:ext} (black dashed curve in the bottom panel). However, the extinction is not grey over this range of wavelengths, ranging from $\tau_{eff} \sim 2.5$  at 0.5 \mic\, to $\sim 0.7-1.5$  at 2.3 \mic.

Given the complexity of the dust in the immediate stellar environment and that of its geometry, it is impossible to  fully constrain the dust properties from the data shown in   Fig.~\ref{fig:ext}.  In particular, the actual contribution of scattering to the total extinction depends on a number of parameters, such as the geometry of the dusty layer and the grain properties, namely the phase function. However, some considerations can be of interest. In the following, we only consider silicate grains,  because, as mentioned before,  there are indications that the dimming layer is close to the star, where carbonaceous materials are likely to evaporate \citep[e.g.,][and Sect. \ref{sec:IRexcess}]{Lodders2003}. In particular, we used  olivine 
(Mg$_2$SiO$_4$), whose optical properties we obtained from the Jena database, \citet{Jaeger2003_Jena}, assuming spherical grains with a 30\% porosity. Scattering and absorption efficiency were computed with the Mie theory and the Bruggeman effective medium description of porosity \citep{Krugel_book}. We adopted a power-law grain size distribution with an exponent 3.0 between a minimum size $a_{min}=0.06  \mu$m and a maximum $a_{max}$. The exact value of the porosity is not important, as long as it is not extreme. 
For  $a_{max}=0.1$ \mic\ the dependence on wavelength is very steep, comparable to that of the ISM (blue solid line in the bottom panel of Fig. \ref{fig:ext}). To reproduce the observed trend of $\tau_{eff}$ with  wavelength, it is necessary to increase $a_{max}$ to much higher values. Such large grains have very high albedo ($\sim 1$ in our range of wavelengths) but strongly  forward-peaking scattering, which in practice reduces the albedo by large factors  \citep[e.g.,][]{Krugel_book,Krugel2009,Mulders2013}. We find good agreement with the observations for $a_{max} \sim 150$ \mic, 
$Q_{tot}=Q_{\rm abs} +0.12 Q_{\rm sca}$ (green curve in the bottom panel of Fig. \ref{fig:ext}); the effective albedo is $\sim 0.7$  in the visual range, consistent with the polarimetry results analysed most recently by \citet{Dodin2019}. 
If the scattering is less efficient, $a_{max}$ needs to be larger. On the other hand, the value of $a_{max}$  can be reduced if grains with different optical constants are adopted. For example, the "astronomical silicates" of \citet{Draine1984}, can reproduce the effective optical depth of \rw for a size $a_{max} \sim 15$ \mic. 


A detailed discussion of the dust properties requires a number of assumptions on the location and extension of the dimming layer and  detailed radiation transfer calculations that include non-isotropic scattering; this is beyond the scope of this paper. However, we  have not been able to find any type of dust that does not require grains much larger than those in the ISM.

\subsection{Infrared excess}
\label{sec:IRexcess}

\begin{figure}[ht!]
\centering
\includegraphics[width=\columnwidth]{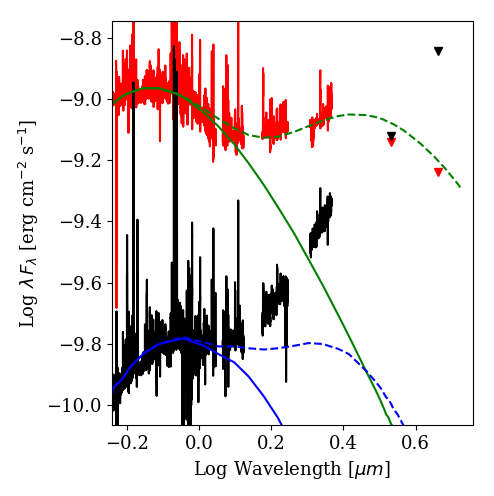}
\caption{X-Shooter spectral energy distribution of the bright state is shown in red.  The red triangles show the WISE flux obtained in March 2014, when RW Aur A was in a  bright state, assuming that half of the flux is contributed by RW Aur B, which is included in the WISE beam. The green curves show the photosphere, approximated by a black-body with $T=4900$\,K and $L=1 L_\odot$, and the total emission of the photosphere plus an additional black-body at $T=1180$\,K and area=$\sim 5.6\times10^{24}$\,cm$^2$. The X-Shooter spectrum in the dim state is shown in black. The black triangles are the WISE fluxes, computed assuming that RW Aur B does not vary during the period of the observations. The blue curves show the photosphere and  the best-fitting model of the bright state emission reddened according to the extinction curve derived in Sect.~\ref{extinction}.}
\label{fig:seds}
\end{figure}

In its bright state, RW Aur A has a very strong NIR excess, as shown in Fig. \ref{fig:seds}. After subtracting the photospheric emission, the excess luminosity in the interval 1.1-2.4\,\mic\ is 0.17 \Lsun, about 5\%-15\% of the stellar luminosity (see Table \ref{tab:param}). In the K band, the observed flux value is three times the photospheric flux. A disc model with $\mathrm{\dot{M}_{acc}\sim 1.0\times 10^{-8}}$ \msun/yr (see sect. \ref{accretion}) can only account for 30\% of the photospheric flux at most.  A very puffed-up inner rim can increase the NIR excess emission \citep{Dullemond2003,Mcclure2013}. The properties of the emitting region can be approximated by fitting a black-body of temperature \Tbb\, to the observed spectral energy distribution.
We find a good fit for \Tbb =1180\,K,  and a projected area of the emitting region  $\sim 5.6\times10^{24}$\,cm$^2$.
The corresponding luminosity ($\sigma T_{\rm BB}^4 \times $Area) is 0.16 \Lsun.
Using eq.(14) of \citet{Dullemond2001},
we  estimate that the distance from the star where  $T=1180$\,K is $\sim 0.1$ au.  Assuming that the emitting region is a cylinder of radius 0.1 au,  we derive a  height  $H\sim 0.12$ au (Area=$2 R \times H$). The ratio $H/R \sim 1$ is  not consistent with  rim properties if hydrostatic equilibrium holds and suggests that even in its bright state \rw is surrounded by an optically thick cloud of  dust close to the sublimation temperature that covers a large fraction of 4$\pi$.   

In the dim state, we firstly assumed that the observed spectrum  is that of the bright state, extinguished as discussed in Sec.~\ref{extinction}. When we adopt the  $\tau_{\rm eff}$ that fits the CO bandheads, then 
 the NIR excess in the dim state is high and increases steeply with wavelength. 
 
 A confirmation of this is  provided by the  WISE archival data at 3.4 and 4.6\,\mic\ discussed
by \citet{Bozhinova2016}. The WISE observations were obtained in one to two consecutive nights in March 2014, when RW Aur A was in a bright state, and in July 2014, February 2015 and July 2015, when the star was in a faint state. Each period has magnitude variations of 0.5-1 mag, but there is  clear evidence that while the 3.4\,\mic\ luminosity does not vary significantly between the two states, the system is much brighter at 4.6\,\mic\ during the dimming than in its bright state \citep[see also][]{Shennavrin2015}.  We show the WISE photometry in Fig.\ref{fig:seds}. Because the WISE photometry includes both RW Aur A and RW Aur B in the beam, we conservatively assumed that both components have similar fluxes in the bright state and that RW Aur B does not vary when RW Aur A enters the dimmed phase. The trends are similar when we assume that the contribution of RW Aur B is negligible in all cases. Moreover, we recall that the WISE and X-Shooter observations are not simultaneous.

In spite of these uncertainties, there is strong evidence that in the dim phase there is not only absorption of the radiation of the innermost region of RW Aur A, but also an additional, warm emission, requiring grains with a range of temperatures of between 500-700 K. This is in agreement with \citet{Shennavrin2015}.

\subsection{Accretion luminosities and mass accretion rate}
\label{accretion}
Knowledge of the extinction during the dimming event with good wavelength resolution allows us to measure the accretion luminosity when the dimming occurs and compare it with the luminosity in the bright state.

\begin{figure*}[ht!]
\centering\includegraphics[width=\columnwidth]{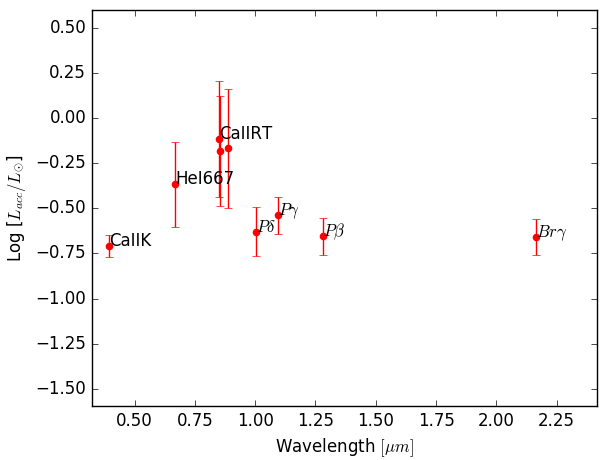}
\includegraphics[width=\columnwidth]{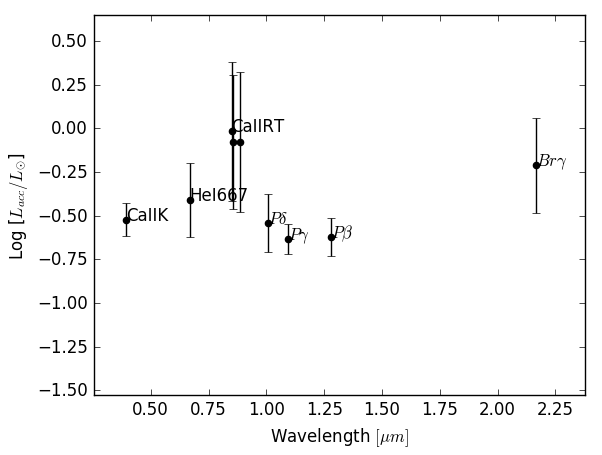}
\caption{Left: Accretion luminosities versus wavelength for the bright state. The fluxes of the optical lines have been corrected for Av=0.4 \citep{White2001}. Right: Accretion luminosities versus wavelength for the dim state. The fluxes have been corrected for the observed extinction curve of Fig. \ref{fig:ext}.}
\label{fig:acc_lum}
\end{figure*}

 The accretion luminosity can be derived with good accuracy from the luminosity of many different emission lines \citep[see, e.g.,][and references therein]{Calvet2004,Alcala2014,Alcala2017}. In RW Aur A, the main difficulty arises from the fact that many lines have very complex profiles, with numerous, deep absorption features overlaying the emission. 
We selected the set of lines in the bright and the dim spectrum with more regular profiles (e.g. symmetric profiles) and computed their luminosity from the flux-calibrated spectrum by integrating over the continuum-subtracted line profile. The line luminosity was calculated for the distance d=152 pc and converted into an accretion luminosity $L_{\rm acc}$ using the correlations between line and accretion luminosity of \citet{Alcala2014}. The results are shown in the left panel of Fig. \ref{fig:acc_lum} for the bright state, for which we corrected for Av=0.4 \citep{White2001}. For the dim state, we corrected each line for the value of $\tau_{\rm eff}$ derived in Sect. \ref{extinction} at the wavelength of the line, and show the results in the right panel of Fig. \ref{fig:acc_lum}.

The average accretion luminosity is very similar in the two states, $0.40 \pm 0.05\,L_{\odot}$ for the bright
state and $0.51 \pm 0.07\,L_{\odot}$ for the dim state. The corresponding mass accretion rate, computed using the stellar parameters in Table \ref{tab:param}, is $\mathrm{\dot{M}_{acc}\sim 1.7\times 10^{-8}}$ \msun/yr and $\mathrm{\dot{M}_{acc}\sim 2\times 10^{-8}}$ \msun/yr for the bright and dim state, respectively. Therefore, the mass accretion rate of RW Aur A, as estimated above,  does not vary during the dimming. This confirms that this is very likely due to dust extinction.

\section{Discussion}


\subsection {Inner gaseous disc}

Overtone ro-vibrational emission of CO is clearly detected in both spectra. A study of the five bandheads in the bright spectrum allows us to confirm that the CO emission comes from a disc with projected  Keplerian velocity $ \sim 113$ \kms. Depending on the exact value of the inclination, this corresponds to  $\rm v_{k} \sim 120-156$\,\kms\ and a distance from the star of $\sim 0.06$\,to 0.08 au,  within the dust sublimation radius. The model in section \ref{sec:model} of the bandhead emission gives $T\sim 3000\,$K, $N_{\rm CO}\sim 1 \times 10^{21}$\,cm$^{-2}$, consistent with a location in the innermost disc.
This is generally the case for all objects where CO overtone emission has been detected \citep[e.g.][]{Ilee2014}.

The models used in Sec.~\ref{sec:overtone} make a number of simplifying assumptions. 
Some appear well justified a posteriori. The  value of $N_{\rm CO}$ is  much higher than the limit for CO self-shielding \citep[$\sim 10^{15}$\,cm$^{-2}$;][]{VanDishoek1988}. The assumption of  LTE to compute the level population may result in overestimating  $T$ if, for example, UV pumping of the vibrational levels is strong.  Strong UV pumping cannot be ruled out using simple arguments, but it seems unlikely for a relatively cold object with low stellar and accretion luminosity such as RW Aur A. 

The assumption of a single $T$, single $N_{\rm CO}$ emitting ring can also be called into question. The emitting area  required to reproduce the observed CO bandhead flux translates into a narrow ring width $\Delta r/r \sim 0.08$. $T$ and $N_{\rm CO}$ radial gradients in the CO emitting region are very poorly constrained \citep{Ilee2014}, but it is unlikely that significant variations can occur over such a small radial distance.  The fact that the emission is dominated by a hot, narrow ring is confirmed by the location of the features detected in the tail of the first two bandheads, as discussed in Sec. \ref{Results:CO conditions}.



CO  emission in the $\Delta\upsilon=2$ bands is detected in a small fraction of YSOs, ranging from about 17\% in massive, embedded objects (Cooper et al. 2013) to 7\% in HAeBe stars (Ilee et al. 2014). The frequency in lower mass objects is unclear, as most samples are relatively small and biased toward more luminous and/or exotic objects \citep{Carr1989,Chandler1995,Biscaya1997}.  However, there seems to be a tendency to lower detection fraction among lower mass objects. For example, \citet{Ilee2014} detect CO in only 2 of the 53 HAeBE stars in their sample with mass $< 3 M_\odot$. 
We have examined the X-Shooter spectra of about 100 T Tauri stars in Lupus \citep{Alcala2014,Alcala2017} and found that after subtraction of the photospheric spectrum, none shows CO emission. 
Of all the bona-fide CTTS in the literature, CO is detected in emission in only 3 \citep[][this paper]{Carr1989,Eisner2014}, which are listed in Table \ref{tab:TTauri}. 

When detected in the overall sample of YSOs, the CO  luminosity correlates with the \brg\ luminosity \citep{Carr1989,Ilee2014,Pomohaci2017}. However, for a given $L$(\brg), very many CO upper limits are well below the detected values \citep[][]{Ilee2014,Eisner2014,Pomohaci2017}. The correlation has often been interpreted as evidence that accretion and CO emission are physically related because the high temperature and column density required to produce detectable CO emission may occur only when the accretion rate is sufficiently high \citep[e.g.,][]{Carr1989}. 

Nevertheless, it is likely that CO emission is only observed in high-mass TTS with high accretion rates. This seems to be a necessary condition, but given the large number of non-detections, not a sufficient one. \citet{Ilee2014} suggested that the inclination of the system with respect to the observer may play a role because CO detections are confined to objects with moderate-to-high inclination. This could be understood if, for example, the CO emission traces not only the inner disc surface but also an inner disc wall, or the base of a disc wind.  Of the three TTS in Table \ref{tab:TTauri}, the inclination of the outer disc, as measured by ALMA, ranges from 63 deg for DO Tau to only 28 deg for DG Tau \citep{Muzerolle2003,Rigliaco2013}. However, it is possible that the inner disc is misaligned with respect to the outer disc, and that it is much more inclined, as suggested by \citet{Eisner2014} for \rw.

\rw was observed in the past with low resolution by \citet{Carr1989}, who detected CO absorption for $L$(\brg)=$2.2 \times10^{-4}$ \Lsun, peak/continuum ratio $\sim 1.1$. \citet{Eisner2014} did not detect any CO in emission. The \brg\ line with respect to the continuum is much weaker than in our spectra, while its luminosity is comparable. This may indicate that the continuum emission was much brighter to make the CO lines undetectable at that sensitivity. \citet{Beck2008} detected at least the first four CO bandheads in emission. Their Fig.3 shows that the \brg\ peak is about 1.6 times the adjacent continuum. The CO spectrum is similar to what we observe.
It follows that the CO emission from \rw seems to have been remarkably stable over the last ten years.
Exploring CO variability and its correlation with other properties, such as the near-IR continuum, and accretion and wind indicators is a promising way to understand the CO emitting region, and it is certainly worth exploring further.

\subsection{ Dust in the dimming layer}
\label{sec:dust}


The extinction curve derived in Sec.\ref{extinction} suggests that large grains (up to a few hundred microns) exist in the dimming layer. This is consistent with other optical and IR observations \citep{Schneider2015}. Comparing the X-ray absorbing column density and the optical extinction, \citet{gunther2018} found that either the gas-to-dust ratio in the absorber is higher than in the ISM, or the absorber has undergone significant dust evolution. 



In its bright state, RW Aur A has a very strong NIR excess, inconsistent with the emission of a disc or rim in hydrostatic equilibrium. It should be noted that the bright state of our observations occurs after a deep dimming but the excess is quite strong even before the start of the minima \citep{Shennavrin2015}.



\subsection{Origin of the dimming events}

The origin of the dimming events of RW Aur A has been much debated in the past few years. As soon as the dimming events were discovered, several authors interpreted the dimming events as caused by material at $\sim180\,$au from star A being ejected along the line of sight by the dynamical interaction between the discs of the system \citep[e.g.][]{Rodriguez2013,Rodriguez2016,Dai2015}. Support for this theory is given by hydrodynamical simulations, and by the ingress and egress time of the dimming events \citep[see summary in][]{Rodriguez2018}: assuming a Keplerian velocity for the occulting body, ingress and egress times correspond to a distance of $\sim180\,$au.

Recent observational results, together with the spectra shown in this paper, indicate, however, that the dimming events are caused by material that is close to the primary star ($0.1-1$\,au). The first indication is given by the strong NIR excess associated with the dimming events, as clearly shown by the WISE data \citep[][see Fig.~\ref{fig:seds} of this paper]{Bozhinova2016}. In Section \ref{sec:dust} we show that the excess seen during the dimming corresponds to material at $\sim500-700\,$K, within 1\,au from the central star. The surface area of this warm/hot dust component thus increases significantly during dimming, suggesting that the dimming event at shorted wavelengths is due to the same dusty material. A second indication is that the polarisation fraction increases during the dimming event (up to 30\% in $I$ band), showing a clear correlation between $V$ magnitude and polarisation fraction \citep{Lamzin2017}. This indicates that while the star has been occulted, most of the circumstellar material has not, and thus constrains the location of the occulting body. Recent models by \citet{Dodin2019} account for the increase in the polarisation fraction due to scattering by a dust layer in a disc wind. This study agrees with our results and locates the dust screen close to the star. The same argument can be used to explain why the O\,\textsc{i} line at 6300\,\AA\ does not vary in flux between bright and dimmed state \citep[e.g.][]{Takami2016}.

Even though we have reasonable evidence that the dust layer screening the central star is from the very inner region of the circumstellar disc, the physical origin of the high column density of dust is still not clear. At least three hypotheses have been proposed as the physical phenomenon that has led to the dimming events. The first hypothesis is a dust-laden wind \citep[e.g.][]{Petrov2015,Dodin2019}. \citet{Facchini2016} \citep[and earlier][]{France2014} have detected narrow absorption lines at $\sim-60\,$\kms\ in the optical Na and K doublets, and in the N\,\textsc{i}, Si\,\textsc{ii}, and Si\,\textsc{iii} UV lines. The absorption features indicate a disc wind, and the velocities are in agreement with a magnetically driven wind \citep[as photoevaporative flows would present much lower velocities,][]{Facchini2016}. However, mass-loss rates from disc winds are correlated with the mass-accretion rate onto the star, and we do not see a variability in the accretion rate in the dimming events. A possibility would then be that while the gas mass-loss rate does not vary, the dimming events trace strong variations in the dust-to-gas ratio of the disc wind \citep{Garate2018}. This hypothesis is partially supported by the super-solar metallicity of the recently accreted material, as traced by the Fe line seen by $Chandra$ \citep{gunther2018}. From the extinction curve derived in Section~\ref{extinction}, we could constrain the occulting dust to have grains as large as a few hundred microns; connecting the disc wind with the maximum grain size entrained in the flow would be a very interesting dynamical experiment. Models of dust entrainment in magnetohydrodynamic winds indeed suggest that grains as large as $10-100$\,\mic\ can be entrained in winds launched within $<1\,$au from the central star \citep[e.g.][]{Miyake2016}. Finally, a dust-laden disc wind can help explain the strong NIR excess, with a significant part of the solid angle around the central star surrounded by dust. 

Another possibility is that the dimming events are caused by a severe puffing-up of the very inner disc, as proposed for UX Ori objects, that is due to a thermal instability \citep[e.g.,][]{Dullemond2003}. From a simple calculation, the time for a disc to dynamically respond in the vertical direction to a significant change in temperature is $\sim H/c_{\rm s}$, where $c_{\rm s}$ is the local sound speed. Because $H/c_{\rm s}=\Omega$, where $\Omega$ is the Keplerian frequency, the ingress and egress timescale of $\sim30$ days corresponds to a radius of $\sim0.2\,$au (we have assumed a $1.4\,M_\odot$ stellar mass). This result is in agreement with the radius of the dusty screen as derived from the strong NIR excess during the dimming events. The difficulty arising from this simple model is that the puffing-up would need to be extreme. \citet{Rodriguez2018} showed that the inclination of the outer disc obtained from ALMA is $55^\circ\pm0.13^\circ$. Assuming that the disc is co-planar, in order to have the surface of the disc at one 
 scale height across the line of sight would require $H/R\sim0.76$. Some authors have thus suggested that the inner disc could be warped because the inclination of the very inner disc from NIR interferometric measurements is $\sim75^\circ$ \citep{Eisner2014}. However, the structure of the inner disc is likely very different from a simple smooth surface, and the visibility coverage of the interferometric observations is very limited, thus the derivation of the inclination from the NIR can be questioned \citep{Eisner2014}.  

A third explanation links the dimming events to the recent star-disc encounter undergone by the RW Aur system. The hydrodynamical models by \citet{Dai2015} show that the periastron passage occurred $\sim500\,$years ago. \citet{Berdnikov2017} suggest that the interaction that has perturbed the outer disc   has recently arrived at the inner regions, thus causing the dimming events. \citet{Dai2015} have demonstrated that in order to reproduce the morphology of the RW Aur system, the orbital plane of the binary has to be misaligned with respect to the circumprimary disc. The interaction thus generates a bending wave from the outer disc \citep[e.g.][]{Nixon2010}, which then propagates inward with a velocity of $c_{\rm s}/2$ \citep[][]{Papaloizou1995,Facchini2013}. Assuming a temperature of $30\,$K at $10\,$au and a temperature dependence $\propto R^{-1/2}$, half of the sound crossing time (i.e. the time it takes for a bending wave to reach the inner regions) is

\begin{equation}
t = \int_0^{R_{\rm out}} {\frac{2dR}{c_{\rm s}(R)}} \approx 520\ \mathrm{yr},
\end{equation}
where we have taken $R_{\rm out}=58\,$au \citep[as measured from the ALMA observations by][]{Rodriguez2018}. The agreement of the two relevant timescales (periastron passage and sound-crossing time) is suggestive, but not conclusive.

\begin{table*}[h]
\caption{T Tauri stars showing CO emission}             
\label{tab:TTauri}      
\centering                          
\begin{tabular}{c c c c c c c c c c c}        
\hline\hline                 
Object &Mass&Radius&Spec type&$\mathrm{\dot{M}_{acc}}$&jet&NIR excess&inc&\brg&ref \\ 
& \msun&R$_{\odot}$&&\msun/yr&&&$^{\circ}$\\
\hline                        
    \rw&1.4&1.6&K0&$2\times 10^{-8}$&yes&large&50&yes&(1)\\ 
     DG Tau&1.19&3.03&K3-K6V&$7.5\times 10^{-7}$&yes&large&28&yes&(2)\\ 
     DO Tau&0.52&2.31&M0&$5.3\times 10^{-7}$&yes&large&63&yes&(2),(3)\\ 

   \hline                                   
\end{tabular} 
\tablefoot{(1)\, \citet{Rodriguez2013}, (2)\, \citet{Rigliaco2013}, (3)\, \citet{Muzerolle2003}. $\mathrm{\dot{M}_{acc}}$ for \rw is taken from this paper.} 
\end{table*}

\section {Summary and conclusions}

This paper discussed some properties of \rw and its activity, using two X-Shooter spectra, one taken during a deep magnitude minimum in March 2015 and the other about 550 days later (September 2016), when the system was again in its bright state. 

\subsection {Gaseous inner disc}
RW Aur A is one of three T Tauri stars with detected CO overtone emission. We discussed that a high accretion rate seems to be a necessary but not sufficient condition for this emission.

Overtone ro-vibrational emission of CO is clearly detected in both spectra. A study of the five bandheads in the bright spectrum allowed us to determine that the CO emission comes from a disc with projected  Keplerian velocity $ \sim 113$ \kms. Depending on the exact value of the inclination, this corresponds to  $v_k \sim 155-120$\kms\ and a distance from the star of $\sim 0.06$ to 0.08 au, which is within the dust sublimation radius. Simple models of the bandhead emission give $T\sim 3000$\,K, $N(CO)\sim 1\times 10^{21}$\,cm$^{-2}$ and a de-projected area of $2.5-3.1 \times 10^{23}$ cm$^2$, consistent with a location in the innermost disc.
The first two CO bandheads are also clearly detected in the more noisy dim spectrum. The CO profiles are similar in the bright and dim states, and thus the same CO model fit the two spectra equally well. This suggests that the physical conditions of the inner disc do not change when the dimming occurs.

We derived the mass accretion rate from a set of emission lines, excluding those with more irregular profiles, and obtained a value of $\sim 2 \times 10^{-8}$ \msun/yr. We obtained a similar value for the dim state, after correcting the line luminosity for the extinction measured by comparing the two X-Shooter spectra. This is a typical accretion rate in T Tauri stars of $M\sim 1$\msun\, \citep[e.g][]{Ardila2013,Alcala2014,Alcala2017}.

These results indicate that the innermost gaseous disc and the accretion rate do not change much over the years and are basically constant over the interval of time when the deep minima occur.

\subsection {Dimming event}

The results on the inner gaseous disc and mass accretion rate indicate that the deep minima are likely due to a layer of dust obscuring the inner 0.05-0.1 au of the system.
We derived the extinction due to the layer of dust that causes the dimming by dividing the spectrum of the dim and bright states. The results show a rather flat dependence of the effective optical depth on wavelength, ranging from about 2.7 in the blue part of the spectrum ($\sim$0.4 \mic) to 1.2--1.5 in the K band. The ratio of the
CO bandheads gives a value of $\tau_{\rm eff}$ that is higher than in the adjacent continuum by about 0.3 mag. We discussed this result using simple silicate grain models and conclude that grains need to be much larger than in the ISM, with maximum sizes of a few hundred microns at least. This is roughly in agreement with the results of X-ray observations \citep{gunther2018}. 
The extinction curve is very well sampled over a wide range of wavelengths, providing a complete set of data to better understand the properties of dust in circumstellar discs.

RW Aur A has a strong NIR excess in both states. During the dimming, we detect excess emission at wavelengths longer than 2 \mic\ that is due to dust at 500-700 K. The location of this dust at about 0.5 au from the star is not consistent with a location in the outer disc (derived assuming that the ingress time of the minima follows the Keplerian rotation velocity of its location).\\\\

Our results together with data from the literature \citep[e.g][]{Lamzin2017} support the scenario that the dimming events are caused by material that is close to the primary star (within 1 au). Still, the cause of the dimming events is not clear. Several hypotheses have been offered to explain this phenomenon, and we list the more pertinent ones below that are consistent with our results.
\begin{itemize}
\item The dimming events are caused by a dust-laden wind as previously proposed by \citet{Petrov2015}. This is supported by the narrow absorption lines detected by \citet{France2014} and \citet{Facchini2016} and the high infrared excess.
\item A severe puffing-up of the inner disc due to thermal instabilities as in UX Ori objects. This is consistent with the location of the dimming from fitting the spectral energy distribution.
\item A perturbation of the inner disc through tidal interaction with RW Aur B.  According to hydrodynamical simulations of \citet{Dai2015}, periastron occured around 500 years ago, which is consistent with the wave reaching the inner disc recently.
\end{itemize}

\begin{acknowledgements}
We would like to acknowledge the referee, H. Takami, for his constructive comments that helped us to improve the manuscript. M.K. is funded by the Irish Research Council (IRC), grant GOIPG/2016/769 and SFI grant 13/ERC/12907. S.F. and C.F.M. acknowledge an ESO Fellowship. R.G.L has received funding from the European Union's Horizon 2020 research and innovation programme under the Marie Sk\l{}odowska-Curie Grant (agreement No.\ 706320). R.F. acknowledges support from Science Foundation Ireland (grant No. 13/ERC/12907). A.C.G. and T.P.R. have received funding from the European Research Council (ERC) under the European Union's Horizon 2020 research and innovation programme (grant agreement No.\ 743029). A.N. acknowledges the kind hospitality of the DIAS.
\end{acknowledgements}
 \appendix
 \begin{appendix}
\section{Spectra}
 \begin{figure*}[h!]
 \centering
 \includegraphics[scale=0.38]{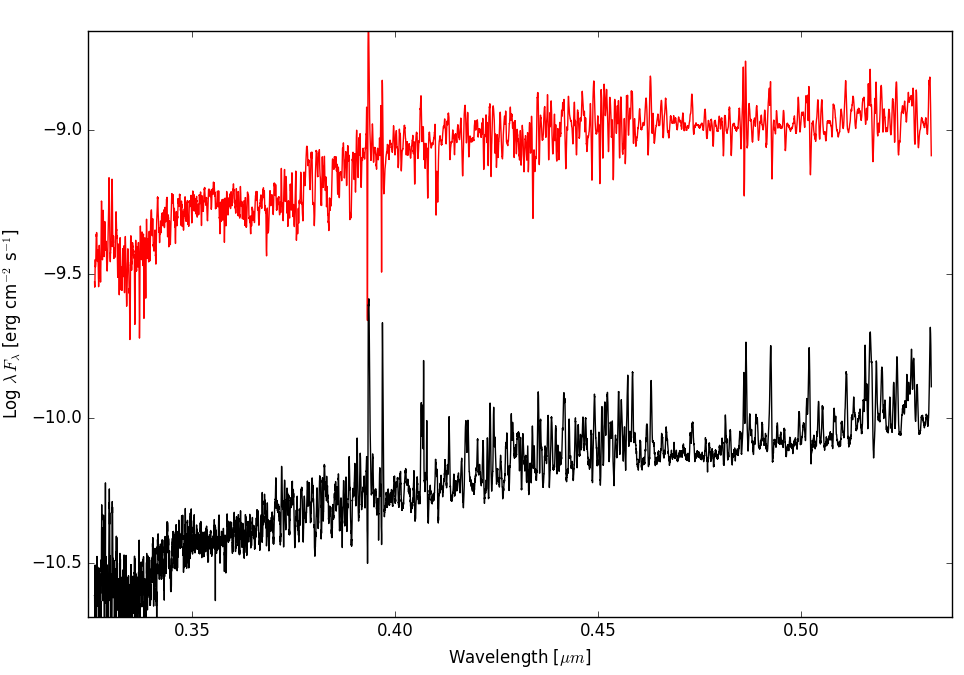}
 \includegraphics[scale=0.38]{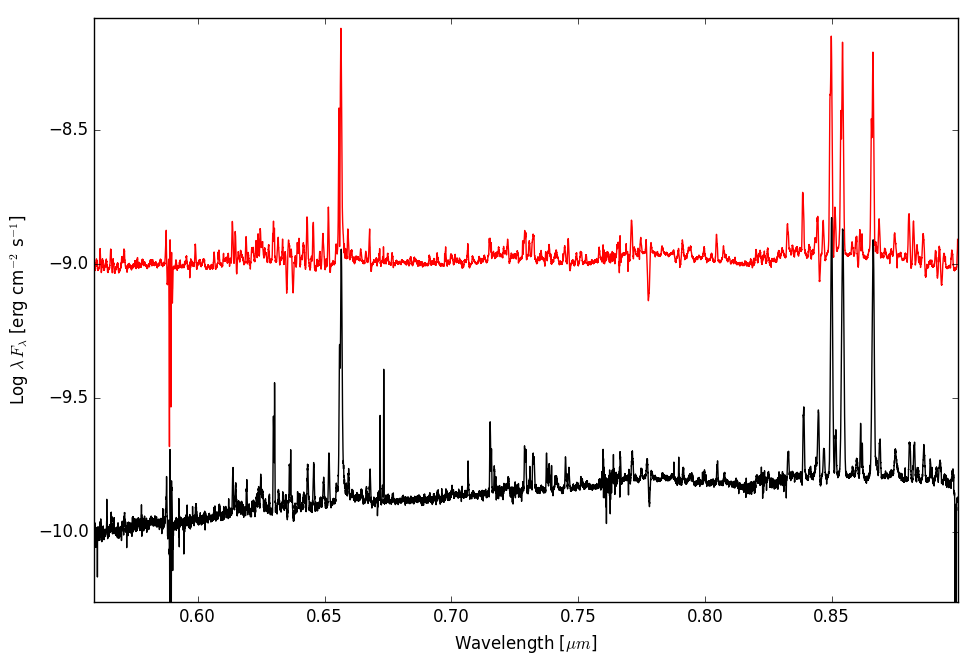}
 \includegraphics[scale=0.38]{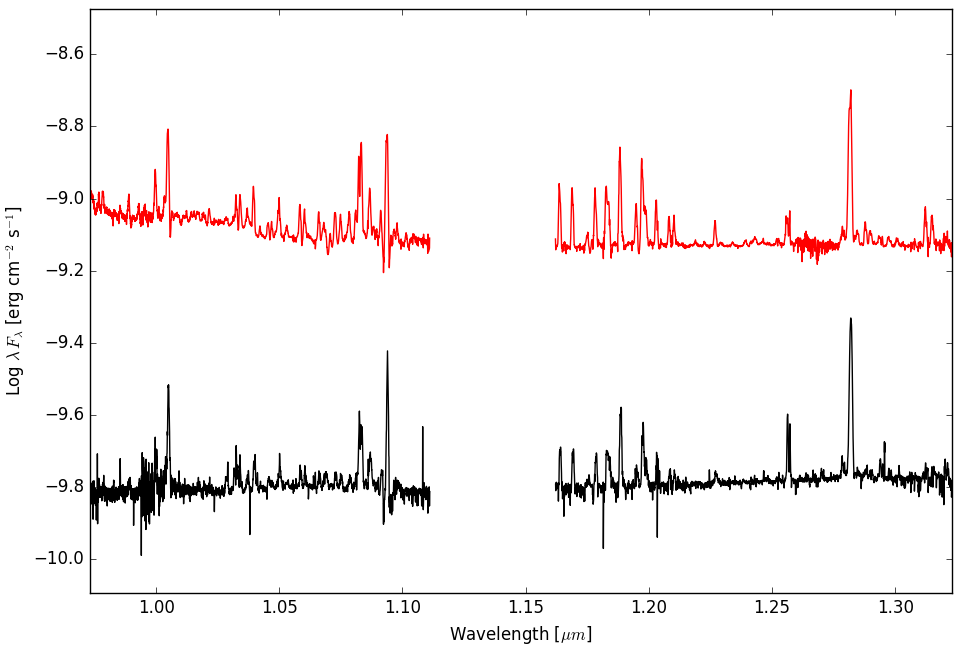}
 \includegraphics[scale=0.38]{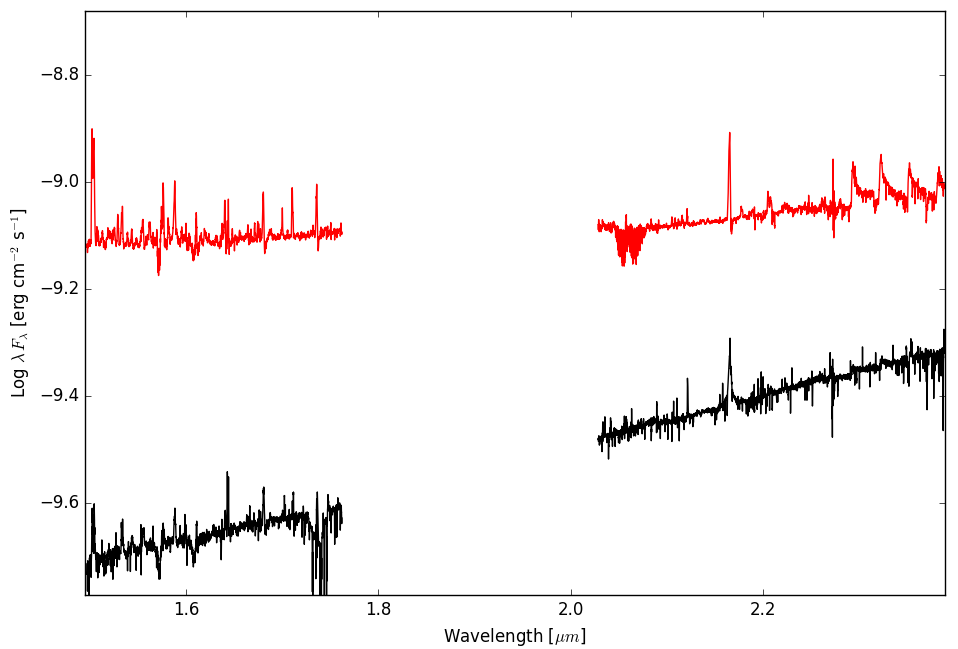}
 \caption{Zoom into the X-Shooter spectra from 0.3-2.4 \mic. The bright state is shown in red and the dim state in black.}
 \label{fig:zoom_spec}
 \end{figure*}

 \begin{figure*}[h!]
 \centering
 \includegraphics[scale=0.7]{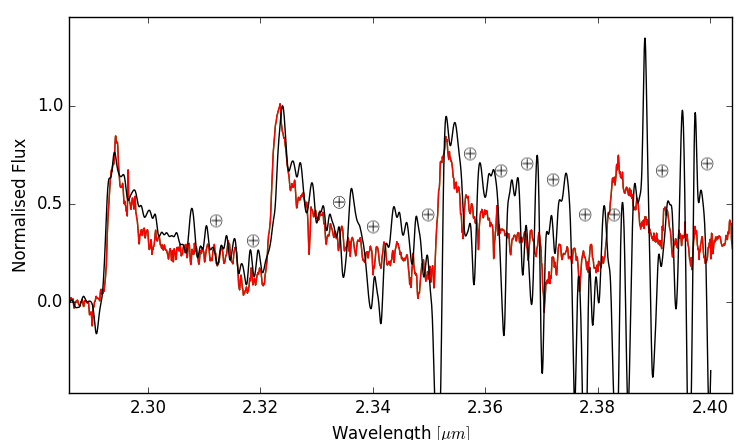}
 \caption{Continuum-subtracted spectra of the CO-emitting regions in the bright (red) and dim (black) states. The spectra are normalised in the second bandhead.}
 \label{}
 \end{figure*}
 \end{appendix}
-----------------------------------------------------
\bibliographystyle{aa}
\bibliography{rw_aur_ref}{}
\end{document}